# DNA Functionalization of Carbon Nanotubes for Ultra-Thin Atomic Layer Deposition of High κ Dielectrics for Nanotube Transistors with 60mV/decade Switching

Yuerui Lu, Sarunya Bangsaruntip, Xinran Wang, Li Zhang, Yoshio Nishi and Hongjie Dai*

*Department of Chemistry and Center for Integrated Systems, Stanford University, Stanford, CA 94305*

Single-walled carbon nanotubes (SWNTs) are advanced quasi-1D materials for future high performance electronics.[1-6] SWNT field effect transistors (FETs) outperform state-of-the-art Si FETs owing to near ballistic electrical transport, chemical robustness, lack of surface dangling bonds and sustained electrical properties when integrated into realistic device structures. One of the goals of transistor down-scaling is to obtain low subthreshold swing $S$ approaching the theoretical limit of 60mV/decade for device operations at low voltages and power dissipations.

For SWNT FETs, vertical scaling of high κ dielectrics by atomic layer deposition (ALD)[7,8] currently stands at $t_{ox}$~8nm with S~70-90 mV/decade at room temperature.[1,3] ALD on free-standing SWNTs is incapable of producing a uniform and conformal dielectric layer due to the lack of functional groups on nanotubes[1,9] and that nucleation of an oxide dielectric layer in the ALD process hinges upon covalent chemisorption on reactive groups on surfaces.[7,8] Existing SWNT FETs with high κ dielectrics rely on ALD nucleation and growth on OH terminated SiO$_2$ substrates and 'spill over' to passively cover SWNTs lying on the SiO$_2$.[1] ALD of relatively thick high κ films (≥8nm) are needed to fully cover the SWNTs and avoid gate leakage. Recently, Gordon and coworkers covalently modified SWNTs by nitro-diazonium to afford conformal ALD coating (10 nm HfO$_2$) on the sp$^3$-modified tubes owing to enhanced nucleation of high κ dielectrics on the nitro-groups on the SWNT sidewalls. Annealing was then used to recover the sp$^2$ structures of the SWNTs and electrical conductance.[9]

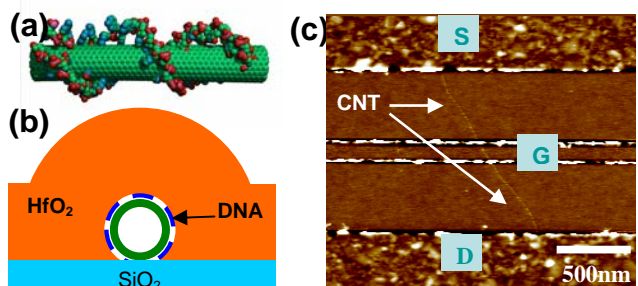

*Figure 1.* DNA functionalization for nanotube electronics. (a) Schematic of a DNA coated SWNT. (b) Cross-sectional view of HfO$_2$ (~ 3 nm by ALD) conformally deposited on a DNA functionalized nanotube lying on a SiO$_2$ substrate. (c) Atomic force microscopy (AFM) image of a high κ SWNT FET with top-gate (G) underlapping source (S) and drain (D).

Here, we show that by non-covalent functionalization of SWNTs with ploy-T DNA molecules (dT40, Fig.1a), one can impart functional groups of sufficient density and stability for uniform and conformal ALD of high κ dielectrics on SWNTs with thickness down to 2-3nm (Fig.1b). This enables us approaching the ultimate vertical scaling limit of nanotube FETs and reliably achieving $S$ ~ 60mV/decade at room temperature. We have also carried out microscopy investigations to understand ALD processes on SWNTs with and without DNA functionalization.

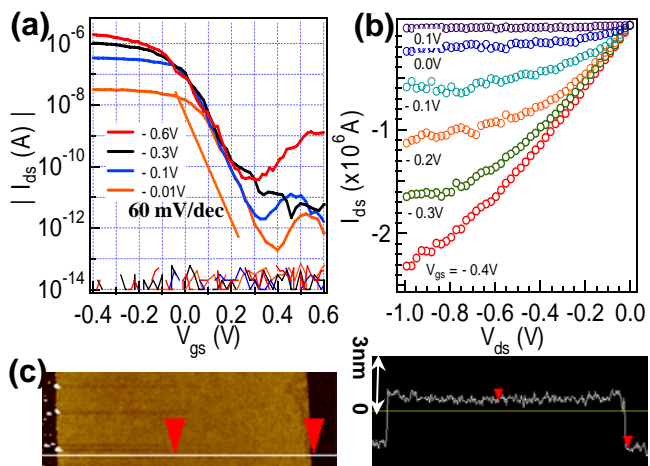

*Figure 2.* A DNA-functionalized SWNT-FET with ~3nm high κ gate dielectrics. (a) Current vs. top-gate voltage ($I_{ds}$-$V_{gs}$) curves of the device in Fig.1c recorded at various bias ($V_{ds}$) indicated. Top-gate leakage currents (bottom traces) under various $V_{ds}$ during $V_{gs}$ sweeps are negligible. (b) Source-drain current-bias ($I_{ds}$-$V_{ds}$) characteristics of the device recorded at various top-gate $V_{gs}$ indicated. (c) Thickness measurement of the HfO$_2$.

Our nanotube FETs were fabricated by patterned growth of SWNTs on 500nm-SiO$_2$/Si substrates,[10] lithographic metal (0.5nmTi/20nmAu contacts) source-drain (S-D) formation, functionalization of the exposed length of SWNTs bridging S/D by soaking the device chip in a 10μM DNA water solution for 30 min followed by 2-min gentle sonication, thorough water rinsing and drying under N$_2$. ALD of HfO$_2$ on the chips at 90ºC using tetrakis-(dimethylamido)-hafnium Hf(NMe$_2$)$_4$ and H$_2$O as precursors[8] and patterning of metal top-gate underlapping[3] the S/D (gate length ~100 nm, Fig. 1c) were then carried out (see Supplementary Information).

Without DNA functionalization, severe gate leakage and shorts were observed for most of SWNT FETs with high κ thickness $t_{ox}$≤5nm. With DNA functionalization, high performance nanotube-high κ FETs free of gate-leakage currents were reliably obtained with HfO$_2$ $t_{ox}$~3nm (Fig.2, diameter of SWNT $d$~1.2nm). All of our $t_{ox}$=3nm DNA functionalized SWNT FETs reproducibly reached theoretical limit of $S$ ~ 60mV/decade at 300K (Fig.2a) and high transconductance of 5000 Sm$^{-1}$ (Fig.2b). For our SWNT FETs with underlapping top-gate, the subthreshold swing $S$ is due to thermally activated on/off switching and $S$= $ln(10)[dV_{gs}/d(lnI_{ds})]=(k_BT/e)ln(10)(1+\alpha)$, where $k_B$ is the Boltzmann constant, and $\alpha$ depends on various capacitances in the device. The fact that we are reaching the theoretical limit of $S=(k_BT/e)ln(10) \approx 60$ mV/decade at 300K suggests $\alpha$ ~ 0 and that the gate capacitance is much higher than other capacitances in the device. For 3nm HfO$_2$ (Fig.2c), the gate dielectric capacitance per unit length is $C_{OX}$~$2\pi\varepsilon_0\varepsilon/\ln(2t_{ox}/R)$ ~ 6pF/cm, where $\varepsilon$ ~ 25 for HfO$_2$ and $R$ is the radius of the SWNT. This exceeds that of the quantum capacitance[3] $C_{QM}$ ~ 4pF/cm of SWNTs resulted from the finite density of states in the nanotube electronic structure. Thus, with $t_{ox}$=3nm $HfO_2$ gate dielectrics, we are well into a regime where the total gate capacitance of a SWNT FET is dominant by $C_{QM}$. Notably, with non-covalent DNA functionalization, we observed no degradation in the SWNT FET conductance after the functionalization and subsequent ALD.

For SWNTs on $SiO_2$ substrates with DNA functionalization, we observed concurrent ALD of $HfO_2$ on both nanotube and $SiO_2$ surfaces, evidenced by the post-ALD topographic topographic height difference between the nanotube site and surrounding being approximately the tube diameter $d$ (Fig.3a top and middle panels). In contrast, for SWNTs on $SiO_2$ without DNA functionalization, 'grooves' were observed along the nanotubes after ALD of $HfO_2$ (Fig. 3b top and middle panels), indicating the lack of nucleation and growth of $HfO_2$ directly on the tube surface and engulfing of the tube by $HfO_2$ grown on the $SiO_2$. We also carried out ALD on suspended SWNTs grown on transmission electron microscopy (TEM) grids. With DNA functionalization, suspended SWNTs allowed for quasi-continuous and conformal $HfO_2$ coating by ALD (Fig.3a bottom). This differed from 'balling up' of high κ material locally on as-grown tubes presumably on defect site (Fig. 3b bottom).[1,9]

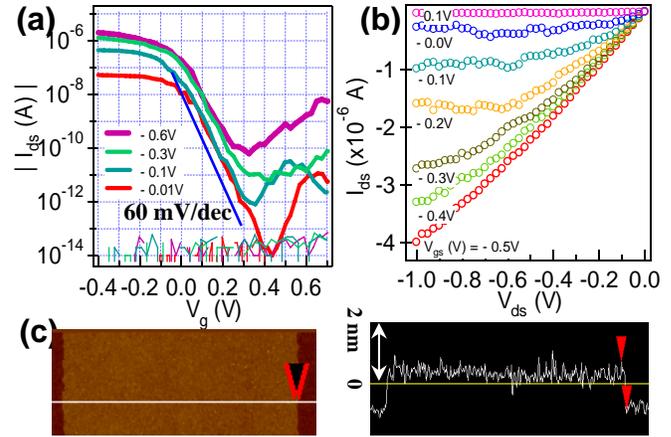

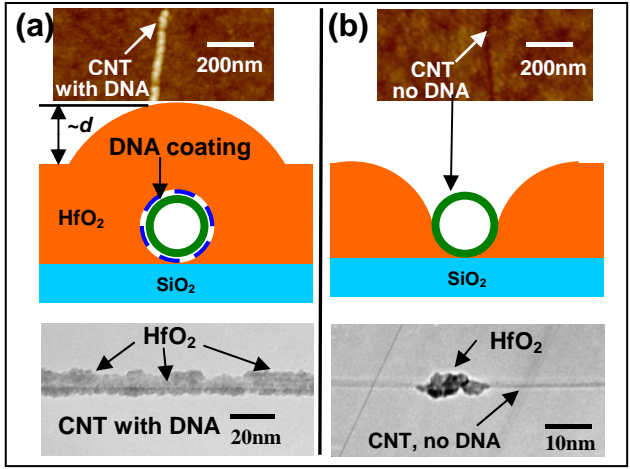

*Figure 3.* (a) ALD of $HfO_2$ coatings on SWNTs with and (b) without DNA functionalization respectively. Top panel: AFM images of ~ 5nm thick $HfO_2$ coatings on SWNTs lying on $SiO_2$. Middle panel: cross-sectional view of the coating profiles. Bottom panel: TEM images of nominally 5nm thick ALD-$HfO_2$ coating on suspended SWNTs.

Our microscopy data provides direct evidence of enhanced nucleation and growth of $HfO_2$ on the sidewalls of DNA functionalized SWNTs in ALD. DNA is known to non-covalently absorb[11] on SWNT sidewalls via π-stacking[12] of the base-pairs. Such functionalization is found here to be thermally stable at the ALD temperature of 90°C. We suggest that the functional groups in DNA molecules (including free OH at the 3'-end and oxygen on the phosphate backbone of DNA) are likely involved in chemisorption of precursor species in the ALD process, affording nucleation sites for the deposition of $HfO_2$ on nanotubes. We have found that functionalization of SWNTs by random-sequence DNA affords less uniform $HfO_2$ coating on SWNTs than dT40 DNA, likely a result of denser packing on SWNTs and higher ability of solubilizing SWNTs in aqueous solutions of the latter.[11]

We attempted pushing the thickness limit of high κ on DNA-functionalized SWNTs and reached a $t_{ox}$~2nm lower limit, below which the uniformity of $HfO_2$ on SWNTs appeared insufficient to avoid gate leakage. For $t_{ox}$~2nm SWNT FETs (Fig.4), we obtained $S$ ~ 60mV/decade and $I_{on}/I_{min}$>$10^4$ at $V_{ds}$=-0.6V (Fig.4a, $I_{min}$ defined as current at the dip of the curve).

Vertical scaling of high κ dielectrics for SWNT FETs below the $t_{ox}$~4-5nm scale affords no further enhancement in transistor switching once $S$~60mV/decade and quantum capacitance are reached. Nevertheless, such scaling is useful for investigating interesting device physics in quasi-1D systems such as electron tunneling. In a 1D channel, electrostatics is dependent on the gate dielectric thickness and widths of tunnel barriers (Schottky, band-to-band tunneling BTBT,[13,14] etc) are often set by $t_{ox}$.[15] Indeed, for SWNT FETs with similar tube diameter of $d$~1.2nm, reduction of $t_{ox}$~3nm (Fig.2) to 2nm (Fig.4) led to increased ambipolar n-channel currents in the latter. The n-channel currents are due to BTBT and a thinner high-κ affords sharper band bending at the edges of the top-gate by the gate potential, thus giving rise to a smaller BTBT width and higher tunnel current. It is proposed recently that BTBT may be utilized to obtain tunnel transistors[14] with $S$ beating the 60mV/decade limit of conventional FETs. The ultra-thin high κ described here should be desirable for such devices. Indeed, under $V_{ds}$=-0.01V, we observe $S$~50mV/decade in the n-channel of our device (Fig. 4a red curve), signaling sharp turn-on of BTBT currents resulted from excellent electrostatic gate control. Thus, non-covalent functionalization can used to enable ultra-thin dielectrics for advanced nanotube electronics.

*Figure 4.* A DNA-functionalized SWNT-FET ($d$~1.2nm) with ~2nm high κ dielectrics. (a)&(b) Device characteristics. Top-gate leakage currents (bottom traces in (a)) under various $V_{ds}$ during $V_{gs}$ sweeps are negligible. (c) AFM image (left) and topography (right) of the ~2nm

**Acknowledgement.** This work was supported by Intel and MARCO MSD Focus Center.

**Supporting Information Available:** Experimental details are available free of charge via the Internet at http://pubs.acs.org.

Table of Contents artwork

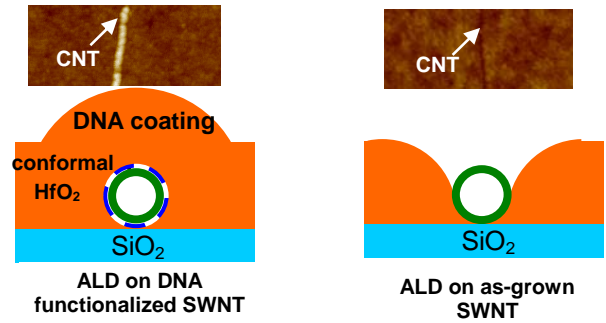

ABSTRACT FOR WEB PUBLICATION

For single-walled carbon nanotube (SWNT) field effect transistors, vertical scaling of high κ dielectrics by atomic layer deposition (ALD) currently stands at ~8nm with subthreshold swing S~70-90 mV/decade at room temperature. ALD on as-grown pristine SWNTs is incapable of producing a uniform and conformal dielectric layer due to the lack of functional groups on nanotubes and that nucleation of an oxide dielectric layer in the ALD process hinges upon covalent chemisorption on reactive groups on surfaces. Here, we show that by non-covalent functionalization of SWNTs with ploy-T DNA molecules (dT40-DNA), one can impart functional groups of sufficient density and stability for uniform and conformal ALD of high κ dielectrics on SWNTs with thickness down to 2-3nm. This enables approaching the ultimate vertical scaling limit of nanotube FETs and reliably achieving S ~ 60mV/decade at room temperature, and $S$~50mV/decade in band to band tunneling regime of ambipolar transport. We have also carried out microscopy investigations to understand ALD processes on SWNTs with and without DNA functionalization.